\documentstyle[12pt]{article}
\newcommand{\gsim}{\mbox{ $
{}_{{}_{\textstyle\sim}}  \! \! \! \! \! {}^{{}_{\textstyle>}}$}}
\newcommand{\lsim}{\mbox{ $
{}_{{}_{\textstyle\sim}}  \! \! \! \! \! {}^{{}_{\textstyle<}}$}}
\begin{document}
\renewcommand{\thefootnote}{\fnsymbol{footnote}}
\thispagestyle{empty}
\vspace*{-1 cm}
\hspace*{\fill}  \mbox{UWITP-94/2} \\
\hspace*{\fill}  \mbox{UWThPh-1994-17} \\
\vspace*{1 cm}
\begin{center}
{\Large \bf Neutralino Mass Bounds
in the Next--To--Minimal Supersymmetric Standard Model
\\ [3 ex] }
{\large F.~Franke\footnote{Supported by
Deutscher Akademischer Austauschdienst
and Cusanuswerk \\ \hspace*{0.5cm}
email: fabian@physik.uni-wuerzburg.d400.de} and  H. Fraas
\\ [2 ex]
Institut f\"ur Theoretische Physik, Universit\"at W\"urzburg \\
D-97074 W\"urzburg, Germany
\\ [3 ex]
A. Bartl
\\ [2 ex]
Institut f\"ur Theoretische Physik, Universit\"at Wien \\
A-1090 Wien, Austria }
\end{center}
\vfill

{\bf Abstract}

We analyze the experimental data from the search for new particles at
LEP 100 and obtain mass bounds for the neutralinos of the
Next--To--Minimal Supersymmetric Standard Model (NMSSM).
We find that
for $\tan\beta \gsim 5.5$ a massless neutralino is still possible,
while the lower mass bound for the second lightest neutralino
corresponds approximately to that for the lightest neutralino
in the Minimal Supersymmetric Standard Model (MSSM).

\vfill
\begin{center}
July 1994
\end{center}

\newpage

\setcounter{page}{1}
\section{Introduction}
The Minimal Supersymmetric Standard Model (MSSM) technically
solves the hierarchy problem of the standard model. However, the
problem remains that one does not know why the
$\mu$ parameter in the superpotential $W \ni \mu H_1 H_2$ should be of the
electroweak scale. Supersymmetric models with an extended Higgs sector
may provide a solution for this $\mu$ problem and, beyond that,
avoid the strong mass bounds of the Higgs bosons
and the neutralinos in the  MSSM.
Models of this kind are also motivated by GUT and string theories
and were first developed in Refs. \cite{barr},
\cite{nilles}, \cite{derendinger}.

The Higgs sector of the MSSM \cite{gunion} contains two
Higgs doublets $H_1$, $H_2$ with vacuum expectation values $v_1$ and $v_2$
($\tan\beta=v_2/v_1$) that lead to five physical Higgs particles,
two CP-even and one CP-odd neutral scalars and a pair of charged
scalars. Furthermore, there are four fermionic neutralino eigenstates being
mixtures of photino, zino and the two neutral higgsinos. In this
simplest supersymmetric model the mass of the lightest neutral
Higgs boson $h$ is bounded at tree level by $m_h \le m_Z|\cos 2\beta |$
(including radiative
corrections this bound can be raised up to 130 GeV
\cite{rad}).
{}From Higgs search at LEP a neutral scalar lighter than
42 GeV and a neutral pseudoscalar with a mass less than 22 GeV
is excluded \cite{l3}. Moreover, LEP data imply that
the lightest neutralino in the
MSSM has a lower mass bound of 20 GeV for $\tan\beta >3$ and the
second lightest is heavier than 40 GeV, a massless neutralino
is only possible for $\tan\beta < 1.7$ \cite{aleph}.

In the
Next-to-Minimal Supersymmetric Standard Model (NMSSM), the minimal extension
of the MSSM by a gauge singlet superfield ${\cal N}=(N,\psi_N,F_N)$ with the
singlet Higgs $N$, the singlet higgsino $\psi_N$ and the auxiliary field
$F_N$, the Higgs sector is extended to five physical neutral
Higgs bosons, three Higgs scalars $S_a$ ($a=1,2,3$)
and two pseudoscalars $P_b$ ($b=1,2$) \cite{higgs}, \cite{drees}.
The singlet higgsino enlarges the neutralino sector to five
neutralinos. The NMSSM is fully determined by fixing the
superpotential and the soft symmetry breaking potential.
The terms relevant for the Higgs sector are
\begin{eqnarray}
W & = & \lambda \varepsilon_{ij} H_1^i H_2^jN-\frac{1}{3}kN^3, \\
V_{\mbox{soft}} & = &  -\lambda A_\lambda \varepsilon_{ij} H_1^i
H_2^j N-\frac{1}{3} k A_k N^2 + \mbox{h.c.},
\end{eqnarray}
where $H_1 = (H_1^0,H^-)$ and $H_2=(H^+,H_2^0)$ are the $SU(2)$ Higgs
doublets with hypercharge $-1/2$ and $1/2$,
respectively, $N$ is the Higgs singlet with hypercharge
0, and $\varepsilon_{ij}$ is totally antisymmetric with
$\varepsilon_{12}=-\varepsilon_{21}=1$.

In the NMSSM the tree level bound on the lightest
Higgs scalar is $m_{S_1}^2 \le m_Z^2 \cos^2 2\beta+
\lambda^2(v_1^2+v_2^2)
\sin^2 2\beta$ \cite{espinosa}, again
it will be raised by radiative corrections \cite{ell}, \cite{elliott}.

Experimental mass bounds from LEP data for the neutral Higgs scalars
were found by
Kim et al.~\cite{kimhiggs}
by studying the production of the lightest Higgs scalar in
$e^+e^- \rightarrow S_1 b\bar{b}$.

Although the parameters $A_\lambda$ and $A_k$ in the soft symmetry
breaking potential strongly determine
the properties of the Higgs bosons, they do not influence
the masses and mixing types of the neutralinos.
Consequently, the Higgs mass bounds of Ref.~\cite{kimhiggs}
do not restrict the other parameters
of the NMSSM
and do not lead to  neutralino mass bounds.
Kim et al.~\cite{kimhiggs}, \cite{kim}, \cite{kimneu}
also calculate cross sections for the production of
neutralinos for special values of the parameters of the NMSSM
but leave the question
open if and in which parameter region a massless neutralino
is still compatible with experimental data.

In this letter we analyze the experimentally excluded domains of the
parameter space and determine neutralino mass bounds in the NMSSM
which originate from experimental data.
The paper is organized as follows: In Sec.~2 we describe shortly the
neutralino sector of the NMSSM. With the experimental constraints presented
in Sec.~3 we then restrict in Sec.~4 the parameter space and
derive in Sec.~5 bounds on the neutralino masses.
Finally, their dependence on the parameters of this model are discussed.

\section{The neutralino sector in the NMSSM}
The NMSSM contains five neutral gauge
and Higgs fermions $\tilde{\gamma}$, $\tilde{Z}$, $\tilde{H}_1^0$,
$\tilde{H}_2^0$, $\tilde{N}$.
The mass eigenstates are the
five neutralinos $\tilde{\chi}_i^0$ ($i=1,\ldots,5$) with
masses and mixings determined by a
$5\times 5$ mass mixing matrix.
The mass matrix depends on six parameters (compared to four
in the MSSM \cite{bartl}): the gaugino masses $M$ and $M'$, the
ratio of the vacuum expectation values of the Higgs doublets
$\tan\beta=v_2/v_1$, the vacuum expectation value of the Higgs
singlet $x$, and the trilinear couplings in the superpotential
$\lambda$ and $k$.

Taking as basis
the two-component spinors
of the photino, zino, and the neutral higgsinos
\begin{equation}
(\psi^0)^T=(-i\lambda_\gamma,-i\lambda_Z,\psi_H^a,\psi_H^b,\psi_N)
\end{equation}
with
\begin{eqnarray}
\psi_H^a & = & \psi_{H_1}^1\cos\beta-\psi_{H_2}^2 \sin\beta ,\\
\psi_H^b & = & \psi_{H_1}^1\sin\beta+\psi_{H_2}^2 \cos\beta ,
\end{eqnarray}
the neutralino mass terms
in the Lagrangian read
\begin{equation}
{\cal L} = -\frac{1}{2} (\psi^0)^T Y \psi^0 + \mbox{h.c.}.
\end{equation}
The mass matrix
\begin{equation}
Y= \left(
\begin{array}{ccccc}
-M s^2_W-M' c^2_W &
(M'-M) s_W c_W &
0 & 0 & 0 \\
(M'-M) s_W c_W &
-M c^2_W-M' s^2_W &
m_Z & 0 & 0 \\
0 & m_Z &
-\lambda x \sin2\beta &
\lambda x \cos2\beta &
0 \\
0 & 0 &
\lambda x \cos2\beta &
\lambda x \sin2\beta &
\lambda v \\
0 & 0 & 0 &
\lambda v &
-2kx
\end{array}
\right).
\end{equation}
with
\begin{equation}
s_W \equiv \sin\theta_W, \hspace{1cm}
c_W \equiv \cos\theta_W, \hspace{1cm}
v \equiv \sqrt{v_1^2+v_2^2}
\end{equation}
can be diagonalized by a unitary $5\times 5$ matrix
$N$
\begin{equation}
N_{im} Y_{mn} N_{kn} = \delta_{ij} m_{\tilde{\chi}_i^0},
\end{equation}
where $m_{\tilde{\chi}_i^0}$ is the mass eigenvalue of the
neutralino state
\begin{equation}
\tilde{\chi}_i^0 = \left( \begin{array}{c}
\chi_i^0 \\ \bar{\chi}_i^0 \end{array} \right)
\hspace*{1ex} , \hspace*{1cm}
\chi_i^0 = N_{ij} \psi^0_j .
\end{equation}
Note that with $\mu=\lambda x$ the upper $4\times 4$ matrix reproduces
the neutralino mass matrix of the MSSM, and the chargino sector
is recovered.

Due to theoretical considerations we restrict the parameter space
in the following way:
\begin{enumerate}
\item In order to avoid explicit
CP violation in the scalar sector we choose
$\lambda k >0$ \cite{higgs}.
Because of the symmetries of the neutralino eigenvalue
equation it is sufficient to consider
$\lambda, k >0$.
\item The vacuum state can be chosen such that
$v_1,v_2,x > 0$
\cite{higgs},
\cite{pietroni}.
\item The assumption of grand unification implies relations
between the gaugino masses $M$ and $M'$ as well as between
$M$ and the gluino mass $m_{\tilde{g}}$.
At the electroweak scale one expects
\begin{equation}
\label{mms}
M'  =  \frac{5}{3} \frac{\alpha_1}{\alpha_2} M \simeq 0.5 M
\hspace{1ex} , \hspace{2ex}
|M|  =  \frac{\alpha_2}{\alpha_3}m_{\tilde{g}} \simeq 0.3
m_{\tilde{g}} ,
\label{mgluino}
\end{equation}
where $\alpha _{i}=g_{i}^{2}/(4\pi),i=1,2,3$ and the $g_i$ are the gauge
couplings of the $U(1)_Y$, $SU(2)_L$ and $SU(3)_C$, respectively.
\item As suggested by naturalness arguments, by the hierarchy
problem \cite{hier}, and by fine tuning constraints \cite{fine},
the gluino mass is assumed to be not much larger than
\mbox{1 TeV}.
\end{enumerate}
Note that the sign of the gaugino mass parameter $M$, however,
remains arbitrary.

In the following, we obey the unitarity bounds on the couplings in the
Higgs potential as worked out in \cite{durand}, but we do not restrict
ourselves to special solutions of renormalization group equations
with fixed boundary conditions
as discussed e.~g.~in \cite{higgs} and \cite{ellwanger}.
We will not impose cosmological constraints obtained by assuming
that the lightest neutralino is the main component of dark matter.
As shown in \cite{dark}, the lightest supersymmetric particle (LSP)
of the NMSSM is most likely expected to lie in the mass range
between 10 and 60 GeV in order to give a sufficient contribution
to the relic density. This requirement would mean that the lower
bounds on the mass of the lightest neutralino as derived in Sec.~5
have to be increased.

\section{Experimental constraints}
The parameter
space of the NMSSM and the masses of the supersymmetric particles
are constrained by the LEP results for
\begin{enumerate}
\item the upper limit of new physics contributing to the
total $Z$ width.
The L3 Collaboration obtained \cite{l3}
\begin{equation}
\Delta \Gamma _Z \leq 35.1 \; \mbox{MeV}.
\label{total}
\end{equation}
\item the upper limit of the contribution of new physics to the
invisible $Z$ width
\cite{l3}
\begin{equation}
\Delta \Gamma _{\mbox{inv}} \leq 16.2 \; \mbox{MeV}.
\label{invisible}
\end{equation}
\item limits from unsuccesful direct neutralino search
in $Z$ decays at LEP.
The ALEPH Collaboration performed a detailed analysis of
the search for the second lightest neutralino within the
framework of the MSSM \cite{aleph}. Taking into
account the decay channels
$\tilde{\chi}^0_2 \rightarrow \tilde{\chi}^0_1 Z^{\ast}
\rightarrow \tilde{\chi}^0_1 f \bar{f}$,
$\tilde{\chi}^0_2 \rightarrow \tilde{\chi}_1^0 \gamma$ (via
loop diagrams) as well as
$\tilde{\chi}^0_2 \rightarrow \tilde{\chi}^0_1 H^0$
we extract for the
branching ratios of $Z$ decays into two neutralinos (not
$\tilde{\chi}^0_1 \tilde{\chi}^0_1$)
\begin{equation}
B (Z \rightarrow \tilde{\chi}^0_i \tilde{\chi}^0_j)
< 5 \times 10 ^{-5} \hspace*{0.5cm} (i,j)\neq (1,1).
\label{direct}
\end{equation}
\end{enumerate}
For special parameters of the MSSM the CDF Collaboration found a
lower bound for the gluino mass $m_{\tilde{g}}$
(in a SUSY scenario with cascade decays of the gluino)
\cite{cdf}
\begin{equation}
\label{cdfbound}
m_{\tilde{g}} > 100 \: \mbox{GeV}.
\end{equation}
By eq.(\ref{mgluino}) this corresponds to
\begin{equation}
M > 30 \; \mbox{GeV}.
\label{gluino}
\end{equation}
Since in the NMSSM more cascade
decays may be possible this limit could even be lower.
Therefore we will not generally restrict the parameter $M$
by eq.~(\ref{cdfbound}), but will dicuss its consequences
in connection with Fig.~4.

\section{Constraints on the parameter space in the NMSSM}
We first show for various values of the parameters
$\lambda$, $k$ and $\tan\beta$ the domain
in the $M-x$ plane excluded by the
experimental bounds presented in Sec.~3.
They are determined by calculating for
every set of parameters the masses and eigenstates of the neutralinos
and charginos
and, with
the couplings
given in \cite{neuprod}, the $Z$ decay width.
Since the $Z$ boson does not couple to the singlet higgsino,
the differences between the MSSM and
the NMSSM just arise by the explicit form of the diagonalization matrix
$N_{ij}$.
If the result violates one of the experimental constraints
this parameter set is excluded.

For $\lambda=0.5$, $k=0.2$ and $\lambda=0.2$, $k=0.005$ the
$M-x$ parameter space excluded by measurements of the total $Z$ width
eq.~(\ref{total}) and direct neutralino search eq.~(\ref{direct})
is depicted in Fig.~1 for $\tan\beta=2$ and $20$, respectively.
The limit on the invisible $Z$ width eq.~(\ref{invisible}) does
not further constrain the allowed region for the considered
parameters.

While the shape of the excluded domain in the $M-x$ plane of the NMSSM
resembles that in the $M-\mu$ plane of the MSSM \cite{aleph},
there are, however, some fundamental differences.
Generally, the allowed parameter space shrinks for increasing values
of $\tan\beta$ and with decreasing parameters $\lambda$ and $k$.

For all parameter values considered there exists an excluded
$M$-region, and therefore by eq.~(\ref{gluino}) also gluino mass
bounds are imposed by LEP data. For
$\tan\beta\lsim 4$ and $k\gsim 0.01$ a light gluino scenario even with
$M=0$ GeV, which also leads to a massless photino, is allowed
in the range $x\le 1000$ GeV while a certain domain of positive $M$ values
is  always excluded.
In our example in Fig.~1 with $\lambda=0.5$, $k=0.2$, $\tan\beta=2$
negative $M$-values are not restricted, but positive values
$41$ GeV $<M<57$ GeV are excluded.

For $\tan\beta \gsim 4$ the LEP data entail a lower bound for $|M|$
of about $45$ GeV corresponding to
$m_{\tilde{g}}>145$ GeV;
the exact bound depends on the maximal
$x$ value considered.

\section{Neutralino mass bounds in the NMSSM}
In this section we
present bounds on the masses
of the NMSSM neutralinos and discuss their dependence on the
model parameters $M$, $x$, and $\tan\beta$.
Our procedure is similar to that described in Sec.~4: We fix some parameters
($\tan\beta$, $M$ and $x$ in the case of Fig.~2, $\tan\beta$
and $M$ for Fig.~3 and $\tan\beta$ for Fig.~4) and vary the
remaining parameters over the range
$-\: 400 \; \mbox{GeV} \le M \le +\: 400 \: \mbox{GeV}$, $0 \le x \le 1
\; \mbox{TeV}$,
$0 \le \lambda,k \le 1$.
Within the allowed parameter
region we then search for maxima and minima of the neutralino masses.

Fig.~2 shows the lower and upper bounds on the neutralino masses
as a function of
the singlet vacuum expectation value $x$ for $M=200$ GeV and
$\tan\beta=2$ and $20$, respectively.
Although the bounds look quite similar for both values of
$\tan\beta$ there is a fundamental
difference: for $\tan\beta=20$ the lightest neutralino could
be massless, while for $\tan\beta=2$ the mass of
$\tilde{\chi}_1^0$ has a lower limit which sometimes turns out to
be very small. The precise dependence on
$\tan\beta$ will be discussed in Fig.~4.
Moreover, there exists a lower limit for the
$x$-values being compatible with the LEP data. Whereas for $M=200$ GeV
$x$ must be larger than approximately 75 GeV,
smaller values of $x$ are allowed
for decreasing values of $M$. In the vicinity of this limit, the allowed
neutralino mass spectrum is rather restricted, for increasing $x$
it becomes broader. For larger $x$-values ($x > M$) the
dependence on $x$ of the lower
mass bounds for all neutralinos is rather weak. The upper bounds
are determined by the asymptotical behavior of the mass eigenvalues:
for the two lighter neutralinos it is almost independent on the $x$-values
(asymptotically
$m_{\tilde{\chi}_1^0} = M'$ and
$m_{\tilde{\chi}_2^0} = M$),
while for the heavier neutralinos it is
approximately
$2kx$ for
$m_{\tilde{\chi}_5^0} $ and $\lambda x$ for
$m_{\tilde{\chi}_3}$ and
$m_{\tilde{\chi}_4}$.

In Fig.~3 we show the dependence of the neutralino mass bounds
on the gaugino mass parameter $M$. As discussed in Sec.~3, for
$\tan\beta=2$ the region $44$ GeV $<M<52$ GeV and for $\tan\beta=20$
the region $-45$ GeV $<M<46$ GeV is excluded.
Apart from $M=0$ GeV, for $\tan\beta=20$
a massless neutralino could exist only for
positive $M$,
while for $M<-5$ GeV and $\tan\beta=20$ there exists a lower mass bound of
1.5 GeV. Similarly, for $\tan\beta=2$
a massless eigenstate
is excluded except for $M=0$ GeV, and
we found a lower bound of 2 GeV for $|M|>5$ GeV.
The lower limit on $m_{\tilde{\chi}_2^0}$ is nearly
constant over a wide range of $M$ values, it just decreases for small
values of $M$.
All upper limits have the same asymptotic behaviour as discussed
above (the horizontal lines in Fig.~3 b) and d) correspond to
the values $2kx$ and $\lambda x$, respectively, for the highest
values of $x$, $\lambda$ and $k$ considered).

Finally, we present in Fig.4 the lower mass bounds of the lightest and
second lightest neutralino as a function of
$\tan\beta$ for $M>30$ GeV. Taking into account only the LEP constraints and
allowing small $M$ values, a massless neutralino would be possible
for all values of $\tan\beta$. Therefore we assume in Fig.~4 that
a lower bound
for $M$ based on the CDF data
according to eq.~(\ref{cdfbound}) also holds in the
NMSSM.

In contrast to the MSSM, the lower mass bounds then decrease with
increasing $\tan\beta$, and for $\tan\beta \gsim 5.5$
a massless neutralino becomes compatible
with the experimental data. For all values
of $\tan\beta$, the mass of the LSP
$m_{\tilde{\chi}^0_1}$ could be as small as a few GeV.
Comparing our results with those of the ALEPH collaboration for the
MSSM \cite{aleph} we find that the mass bounds for the
second lightest neutralino
in the NMSSM correspond nearly to those for the lightest one in the MSSM,
and so on for the next neutralinos, while the lightest neutralino
of the NMSSM can be much lighter than in the MSSM or even massless
for $\tan\beta \gsim 5.5$. Therefore in the NMSSM
the neutralino spectrum is similar to that of the
MSSM with the exception of an additional very light or even massless
singlet like neutralino. This could lead to interesting phenomena
for the production and decay of neutralinos at
the next generation of high energy colliders.

\newpage
\section*{Figure captions}
\begin{tabular}{lp{11cm}}
Figure 1: &
The excluded parameter space in the $M$--$x$ plane for
various values of $\lambda$, $k$ and $\tan\beta$:
from total $Z$ width
measurements (bright shaded) and direct
neutralino search (dark shaded). \\ & \\ Figure 2: &
Upper and lower bounds on the neutralino masses in the NMSSM
for $M=200$ GeV and
$\tan\beta=2$
(a,b) and $\tan\beta=20$ (c,d). \\ & \\ Figure 3: &
Upper and lower bounds on the neutralino masses
in the NMSSM for
$\tan\beta=2$
(a,b) and $\tan\beta=20$ (c,d). \\ & \\ Figure 4: &
Lower mass bound for the lightest (a) and second lightest (b)
neutralino in the NMSSM.
\end{tabular}
\end{document}